\documentclass[aps,pra,twocolumn,nofootinbib]{revtex4}
\usepackage{graphicx}

\newcommand{\jr}{ }

\begin{document}
\title{Matter-Antimatter Asymmetry - Aspects at Low Energy}
\author{Lorenz Willmann$^{1}$ and Klaus Jungmann$^{1}$} 
\affiliation{$^1$\hbox{University of Groningen, Van Swinderen Institute, Zernikelaan 25, NL 9747AA Groningen} } 

\begin{abstract}
  The apparent dominance of matter over antimatter in our universe is an obvious and puzzling fact which cannot be adequately explained in present physical frameworks that assume matter-antimatter symmetry at the big bang. However, our present knowledge of starting conditions and of known sources of CP violation are both insufficient to explain the observed asymmetry. Therefore ongoing research on matter-antimatter differences is strongly motivated as well as attempts to identify viable new mechanisms that could create the present asymmetry. Here we concentrate on possible precision experiments at low energies towards a resolution of this puzzle.  
\end{abstract}

\maketitle

\section{Introduction}
\begin{figure}

  \includegraphics[width=\columnwidth]{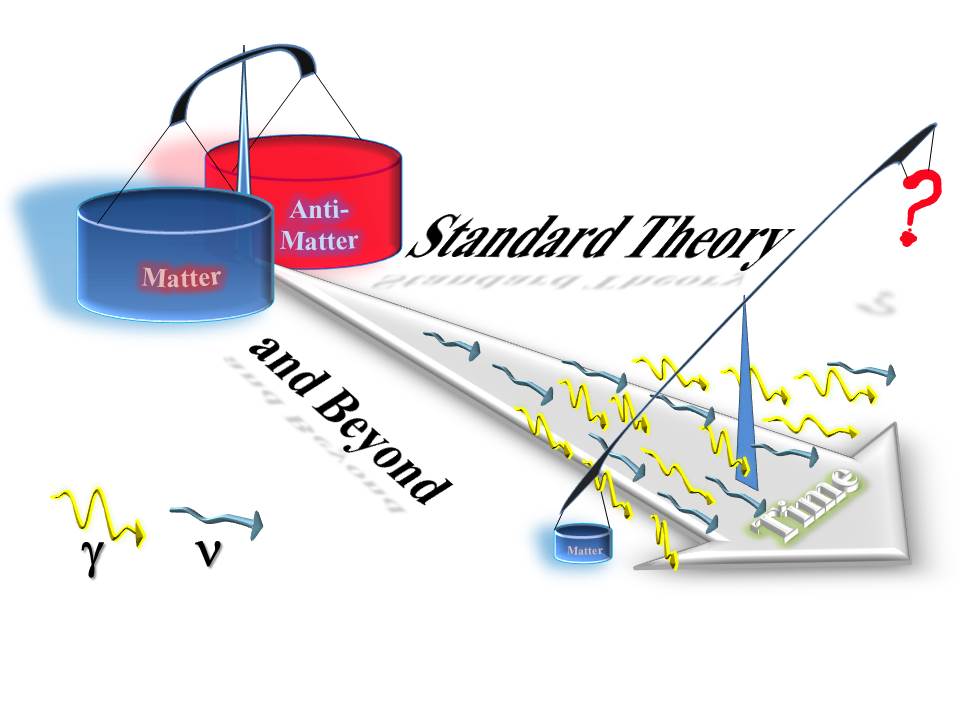}%
   \vspace*{-5mm}
   \caption{\label{fig_Willmann_1}
  In Standard Theory, i.e. the combined Standard Models of particle physics and cosmology, the evolution of the universe matter and antimatter annihilated into photons ($\gamma$) and neutrinos ($\nu$). Today it appears that we have 9 orders of magnitude more photons than baryonic matter particles and no or at least significantly less antimatter. This  asymmetry could also arise at the starting point or later from not yet identified mechanisms.
  }
\end{figure}

Today we have a model of  the universe where objects consisting of matter are surrounded largely by cosmic microwave photon and low energy neutrino background \cite{Weinberg_2006}.
There is evidence for an about 20 times larger amount of dark matter and dark energy \cite{Trimble_1987,Ade_2013}. The numbers of photons 
seems to indicate that most of the primordial matter has annihilated and only a small fraction of matter has survived \cite{Weinberg_2006} (see Fig. \ref{fig_Willmann_1}). The Standard Models of Cosmology \cite{Dolgov_2004} and of Particle Physics \cite{Beringer_2013}, i.e. Standard Theory, contain no known mechanism that would be sufficient to explain this asymmetry.
\footnote[2]{Asymmetric starting conditions such as fluctuations next to the {\it origin of time} causing missing antimatter in our light cone cannot be excluded from first principles. We have insufficient information about this period to be able to exceed speculations. The matter-antimatter puzzle arises in part, because Standard Theory { is built} on assumed symmetries.}
  
Presently we know a model by Sakharov \cite{Sakharov_1967} dating back to the late 1960ies. In this model CP-violation could produce an asymmetry between matter and antimatter, if baryon number $\bf{B}$ is violated and the universe is not in thermal equilibrium. The necessity of the non-equilibrium conditions suggests that the asymmetry at present time would be constant. As of today, the known sources of CP-violation are inadequate to explain observations; next to an experimental proof of $\bf{B}$  violation new sources of CP-violation need to be identified for this model to suffice as a satisfactory explanation. Searches for new CP violation, e.g. in the neutrino sector, are ongoing. 

Another model has been introduced in the late 1990ies \cite{Kostelecky_1997,Glashow_1998},  in which not yet observed CPT-violation could explain the matter-antimatter asymmetry. It has as the only additional condition that baryon number is not conserved. In particular, this model does not require thermal non-equilibrium and therefore the asymmetry could build up continuously in the course of the evolution of the universe. It is, however, a severe drawback of the CPT violation model \cite{Kostelecky_1997,Glashow_1998,Timmermans_2013,Kostelecky_review} that, similar to Sakharov's model, it has no specific predictive power.  By construction, the CPT violation model is perturbative in its nature and it contains infinitely many parameters which relate to Lorentz invariance violation and in some cases also to CPT violation. Values different from zero for many of the parameters have been experimentally searched for in the past 15 years in electromagnetic interactions \cite{Kostelecky_review} with astounding accuracy, and since very recently in weak interactions \cite{Mueller_2013}  without any successful observation, yet. Checking all parameters in a systematic way is an impossible task and unfortunately we have to hope for an accidental finding. 

The theoretical approaches \cite{Sakharov_1967,Kostelecky_1997,Glashow_1998,Timmermans_2013,Kostelecky_review}  have rather limited potential to provide guidance towards specific experimental searches.  They have no defined status as a physical theory, yet.  Their enormous usefulness, however, arises from the fact that they provide a common framework for comparing different experiments which have been conducted so far in various fields. A quantitative comparison in terms of energy scale is enabled. These models yield quantitative information about accessible parameters in potential future experiments, and in particular whether those have been scrutinized before \cite{Kostelecky_review}. Upcoming and numerous possible future experiments at all accessible energy scales would benefit enormously from tighter guidance from out theory. This can help to overcome the present unsatisfactory need for checking physical processes at rather random choice.  
Without future guidance from qualitative and significant upgrades of the models, the fate of presently known approaches towards new CP violation or CPT violation will depend on one or more future serendipitous observations. Therefore, all precise experimental approaches towards CP or CPT violation share a largely equal chance of success. They are all highly motivated as long as the scrutinized parameters have not been limited better by other means before. 

\section{Matter-Antimatter in Experiments}
Searches for differences in properties of particles and corresponding antiparticles in their production and decay rates have been performed. Only rather small CP violation has been established in the hadronic sector of weak interactions (see e.g. \cite{Bigi_2015}), first in the neutral kaon system (see e.g. \cite{Schubert_2015}). Such CP-violation is insufficient to explain the observed asymmetry. 

Another class of experiments explores properties such as electromagnetic moments in precision comparison of particle and antiparticle properties; or such experiments exploit accurate measurements in bound states of particles and antiparticles.
Leptonic atoms like positronium and muonium contain one particle and one antiparticle. They have no defined matter or antimatter status, respectively. They have been subject to precision spectroscopy of atomic gross, fine and hyperfine structure (see e.g. \cite{Hydrogen_atom}) to test QED at the highest precision level. Yet, there is no evidence for any discrepancy. 

Further, a large class of experiments searching for violation of Lorentz invariance also offer a window at processes which can violate CPT and hence could create a difference between matter and antimatter. CPT violation implies Lorentz violation; many possible searches for Lorentz violation also provide for a search for CPT violation, although not all of them. Furthermore, numerous experiments have been conducted on elementary particles and their antiparticles concerning the equality of their parameters; this is a subclass of possible searches for CPT violation.

Although no significant difference could be established yet \cite{Beringer_2013}, these experiments all are strongly motivated and have very high discovery potential. They are one part of the presently ongoing wider class of precise experimental tests of fundamental symmetries \cite{Jungmann_2014}.

\subsection{Fundamental Particle Properties}
In the Standard Model the CPT theorem forces the fundamental properties of particles and antiparticles to be identical. Thus precise studies of properties can set stringent limits on the validity range of the Standard Model, while still having the potential to discover new interactions. Here we discuss selected experimental activities with emphasis on the matter-antimatter asymmetry.

\subsubsection{CP Violation in Electric Dipole Moments}
Permanent electric dipole moments (EDMs) violate both parity (P) and time reversal (T) symmetry.  They are considered a promising route for finding additional sources of CP violation and thereby they might contribute to explain the matter-antimatter asymmetry. The discovery potential of searches for EDMs has led to a plurality of experiments beyond the since long well recognized searches for neutron and atomic EDMs (see e.g. \cite{Jungmann_2013}). 

In this field new systems are being investigated and novel experimental approaches have been developed. Experiments in atoms, in molecules or in ions try to exploit big enhancements of intrinsic elementary particle EDMs in composed systems. Such enhancement can be of order up to $10^6$ for the electron EDM in heteronuclear molecules, e.g. \cite{Commins_1999}. 
 The strongest limit on the electron EDM $d_{e} = 8.7 \times 10^{-29}$ e\,cm (90\% C.L.)\cite{Gabrielse_2013} became possible because of the enhancement in molecules such as YbF\cite{Sauer_2011} and ThO \cite{Gabrielse_2013}. This approach can be expected to deliver further significant improvements over the current limits \cite{Tarbutt_2013}.
A particular important development in this context appears to be the recent experimental progress concerning slowing down of diatomic or even larger molecules \cite{SHoekstra_2014,SHoekstra_2014a} such as, e.g.,  SrF. This enables combining the advantage of large EDM enhancement factors and rather long coherence and observation times. Because of this a further significantly improved sensitivity to an EDM of the electron can be expected.
 
The tightest bound on an atomic EDM comes from $^{199}$Hg at $d_{Hg} = 3.1 \cdot 10^{-29}$ e\,cm (95\% C.L.)\cite{Griffith_2009}. This puts limits on various CP violating mechanisms \cite{Swallow_2013}.
 Nuclear EDMs of Rn and Ra are enhanced due to degenerate levels of opposite parity (see e.g. \cite{Gaffney_2013}) by 3 to 5 orders of magnitude \cite{Flambaum_1999}. This has  stimulated new theoretical and experimental activities in these atoms (see e.g. \cite{Willmann_2009,Parker_etal_2015}).

The control of systematic effects and avoiding misinterpretation of signals is the most urgent issue. 
A particularly interesting possibility arises for spin-precession in gas mixtures of $^3$He and $^{129}$Xe where $^3$He serves as a co-magnetometer occupying exactly the same space as $^{129}$Xe.
The achievable long coherence times and large number of particles promise improvements of up to 4 orders of magnitude over a previous bound  in the $^{129}$Xe atom \cite{Chupp_2001}
and two orders of magnitude over the bound established for $^{199}$Hg \cite{Jungmann_2013,Heil_2013}.  

Another approach towards EDMs 
are light charged particles or nuclei in magnetic storage rings (see, e.g. \cite{Farley_2004,Onderwater_2012,Pretz_2013}). In such an experiment, an EDM could manifest itself as an out of orbit plane precession of the particle spin. In storage rings it is possible to exploit the rather high  motional electric field which a stored particle can experience, when it moves at relativistic velocities in a magnetic field.  Recently there has been good progress in the theoretical understanding of possible mechanisms that could induce EDMs. In particular very light nuclei \cite{deVries_2011} are good candidates for performing an EDM search in a storage ring. However, experiments capable of reaching  $10^{-29}$ e\,cm sensitivity or beyond can not be expected to come online in the near future. They are in a development stage concerning equipment and principal experimental techniques (see e.g. \cite{Brantjes_2012}).

All the EDM experiments have a 
robust discovery potential for new sources of CP violation, which is one of the two yet missing ingredients for Sakharov's model. At the same time they have already a proven record of significantly disfavoring several speculative theoretical models beyond Standard Theory through establishing prohibitive upper bounds on EDMs.

\subsubsection{Antiproton Experiments}
The availability of antiproton facilities has enabled sensitive experiments on this antiparticle and on atomic systems containing antiprotons. Precision measurements on antiprotons and on antiprotonic atoms and ions yield a direct comparison of matter and antimatter particle properties. The first experiments on such exotic atoms were performed in  antiprotonic helium \cite{Hayano_2013}. Together with the theory of bound systems, which comprises all knowledge from matter experiments, properties of the antiproton could be extracted with very high precision \cite{Hori_2011}. They  appear to agree well with the corresponding parameters for the proton, signaling no difference between the two distinct particles. 
The antihydrogen atom plays a unique role since it is the only stable atom synthesized purely from antiparticles. It has been produced \cite{Hbar_production}, trapped \cite{Hbar_trapping} and its charge neutrality was tested at the AD at CERN.  This work is progressing and promises more significant results in the coming years. 

In the first spectroscopy experiment in a pure antiatom hyperfine transitions could be induced in antihydrogen by the ALPHA collaboration\cite{Amole_2013}. The measurement was conducted with magnetically trapped antihydrogen atoms. The positron spin could be flipped with resonant microwave radiation. Affected atoms were ejectect from the trap and could be detected. Because the hyperfine structure arises from a contact interaction between the antiproton and the positron, one can expect that such an experiment would be  more sensitive  to short range new forces than one that measures the atom's gross structure. 
At present several experiments are underway which aim for determining the ground state hyperfine splitting. The ASACUSA collaboration, for example, uses a cold beam of antihydrogen atoms in a Rabi-type atomic beam microwave spectroscopy apparatus \cite{Widmann_2013}. The experiment is progressing fast. A measurement of the hyperfine structure in antihydrogen promises ultimately the best CPT test among the completed and the ongoing experimental efforts

The state-of-the-art of spin flip detection of single protons in the lowest quantum state in a Penning trap \cite{Mooser_2013,DiSciacca_2013} 
opens the path to a measurement of the magnetic moment of particle and antiparticle with this equipment. 
A measurement on single antiprotons by ATRAP yielded a ratio of the magnetic moments of particle and antiparticle 
at 5~ppm\cite{DiSciacca_2013a}.
This, together with the large improvement on the proton\cite{Mooser_2014}, promises a boost for the accuracy of the particle-antiparticle comparison. The recently started  BASE experiment at CERN announced a precision goal  below 1~ppb for the antiproton magnetic moment. 

 The high experimental activity and the significant results in the past decades at the CERN AD lets us expect further progress towards precise comparison of matter and antimatter.

\subsubsection{Muon Anomalous Magnetic Moment}
The measurement of the muon magnetic anomaly $a_{\mu}$ to 0.5 ppm in an experiment at the Brookhaven National Laboratory, Upton, New York, USA can be interpreted as a test of the Standard Model of particle physics. The magnetic anomalies of muons and antimuons agree to 0.7 ppm \cite{Bennett_2004}, while at this time the experimental and the Standard Model based theory value differ at present by some 4 standard deviations. For a decade no issue could be found concerning the experimental value. The theoretical value on the other side underwent in this period several refinements. Presently (2015) the theory values obtained along different routes agree well\cite{theo_g2}.
A new collaboration has been started to measure the magnetic anomaly for  $\mu^+$ at the Fermi National Laboratory, Chicago, USA, with the goal to improve the experimental uncertainty for $a_{\mu}$ by a factor of 5. At this level of sensitivity it could be possible to verify or to disprove a significant difference between theory and experiment. 

The new experiment is presently being installed. It exploits the very same experimental concept, in particular it operates at the so-called {\it magic} momentum, where the influence of motional magnetic fields is canceled. Such motional magnetic fields arise in the muons' rest frame when they are travelling through electric focusing fields in the ring.
The Fermilab experiment reuses crucial parts of instrumentation, in particular the storage ring magnet and the magnetic field 
measurement and control concept. The detectors and the data acquisition system will be upgraded to be able to cope with the expected significantly more intense and cleaner muon beam.
At J-PARC a novel experimental approach \cite{J_PARC_g2} to measure $a_{\mu}$ for  $\mu^+$ is in  its R\&D phase. The experiment employs a small diameter storage ring and operates at lower than the 
{\it magic} momentum. This results in significantly different systematics compared to the
Fermilab experiment. The J-PARC experiment will therefore be very important once the Fermilab approach will have confirmed or not confirmed the present difference between the experimental and the  theoretical values.

\subsubsection{Baryon and Lepton Number Violation}
Global symmetries correspond to conservation laws. In modern physics we know of many conserved quantities where we do not know a fundamental 
symmetry associated with them. Examples are the conservation of  baryon number $\bf{B}$, of lepton number $\bf{L}$ or
of charged lepton flavor  $\bf{L_e}, \bf{L_{\mu}}$ and $\bf{L_{\tau}}$. That calls for precise experiments which search for violations of these
laws. In many speculative Standard Model extensions, global symmetry violations appear naturally. Non-observation 
of such violations can in reverse rule out speculative models. Therefore precision experiments that search
for global symmetry violations have a robust potential to steer model building in fundamental particle physics.

Models employing CP and CPT violation to explain the matter-antimatter discrepancy in the universe require $\bf{B}$  to be violated. All ongoing efforts to find violation of CPT symmetry are therefore most strongly motivated. The present focus rests on searches in large underground laboratories for decays of nucleons, which violate $\bf{B}$  by one unit, and searches for neutron-antineutron oscillations, which violate $\bf{B}$  by two units, in dedicated neutron beam experiments (see e.g. \cite{Babu_2013} for a recent summary of the situation). Both processes probe significantly different physics and at different energy scales. While the Standard Model does not know  $\bf{B}$ as a separately conserved quantity, the difference  $\bf{B}- \bf{L}$ is well conserved in the Standard Model. Therefore also searches for  $\bf{L}$ violation are highly motivated. The observation of neutrino oscillations additionally strongly motivates searches for violation of lepton flavors $\bf{L_e}, \bf{L_{\mu}}$ and $\bf{L_{\tau}}$ among the charged leptons.

Rare muon decays are possible in many speculative theoretical frameworks, including supersymmetry. Such experiments can have rather clean signatures. Some of them have already reached very high precision and new, improved projects are underway.
The MEG collaboration at PSI, Villigen, CH, has recently established a limit on the decay $\mu^+ \rightarrow e^+ + \gamma$. In their COBRA detector, they determine precisely energy and momentum of the monoenergetic $\gamma$ that would be released in the decay. The branching ratio for this decay is below $5.7 \cdot 10^{-13}$ (90\% C.L.)\cite{Adam_2013}. The collaboration expects an improvement by one more order of magnitude.  
Also at PSI a new search for the process $\mu \rightarrow e e e $ has been started. It aims at an ultimate 
sensitivity of below  $10^{-16}$ for the branching ratio \cite{Berger_2014}. The Mu2e
experiment at Fermilab \cite{Fermilab_mue_2013} plans to improve the limit on the process $\mu N \rightarrow e N$. It aims for 
a sensitivity to a branching ratio of order $10^{-17}$. The COMET experiment at J-PARC \cite{JPARC_mue_2013} has a sensitivity goal
of $10^{-18}$ for the same process. It is expected that at such precision, models like supersymmetry can be very decisively tested.

Note, whereas in the quark sector oscillations between $K^0 = (d \overline{s})$ and $\overline{K^0}=(\overline{d} s)$ are well established,  the analogous process of oscillations between the corresponding leptonic bound systems, muonium $M=(\mu^+ e^-)$ and antimuonium $\overline{M}=(\mu^- e^+)$, which both are mixed systems composed of matter and antimatter, have never been experimentally observed in high precision experiments \cite{Willmann_1999,Willmann_1996}.
This leaves an asymmetry between hadronic and leptonic matter which calls for more precise experiments in particular on leptons \cite{Willmann_1996}.

\subsection{Lorentz Invariance}
Within the past two decades numerous precision tests of Lorentz invariance have been conducted. They can be quantitatively 
compared in the framework of the Standard Model Extension (SME) 
\cite{Kostelecky_1997}.
Mind boggling precision { on some parameters} has been achieved. Such experiments test new physics well  beyond the Planck scale. Some of these experiments can provide hints towards 
resolving the matter-antimatter asymmetry.  
Typically, the signature in an experimental search is a spatial anisotropy of 
fundamental properties in selected systems. 
The majority of the experimental efforts were focused on the electromagnetic interaction.
Significant tests in the weak sector - the only sector where violations of discrete symmetries such as C, P, CP and T are known from experiments to exist - have recently been started \cite{Mueller_2013}. Those experiments may 
discover CPT violation and/or Lorentz Invariance violation  already at
much lower precision, since they explore {\it completely new territory}.

The tests of Lorentz invariance in electromagnetic interactions include spin precession of polarized $^3$He and $^{129}$Xe in mixed samples. For spin polarized samples, no correlation of the Larmor precession frequency on the sidereal time could be found. In the SME framework this constrains various Lorentz violating parameters to the level of $10^{-27}$ GeV for protons, 
$3.72\cdot 10^{-34}$ GeV for neutrons and $10^{-31}$ GeV for electrons \cite{Gemmel_2010}. These results have been improved recently by up to a factor of 30 \cite{Allmendinger_2014}.

Lorentz violation in weak interactions is searched for 
in the $\beta$-decay $^{20}{\rm Na} \rightarrow$ $^{20}{\rm Ne} + \beta^+ + \gamma$
at the TRI$\mu$P facility in Groningen, NL. 
The $\beta$-decay rate and asymmetry of a nuclear spin polarized sample has been recorded 
as a function of sidereal time, i.e. the spin orientation in a potentially preferred reference frame. 
A novel limit on Lorentz violation could be established \cite{Mueller_2013,Wilschut_2013} and larger 
recorded dataset promises further improvement by some one order of magnitude. The experiments are 
complemented by theoretical work that enables comparison of various experimentally accessible parameters
in a common framework also for weak processes \cite{Noordmans_2013,Timmermans_2013} .
So far no hint was found for Lorentz or CPT violation.

\subsection{Gravity}
Gravity is not part of the Standard Model. It is the least understood among the four fundamental interactions.
At present there is no theory that provides for combining gravity and quantum mechanics. 
It remains a still unanswered issue, whether the gravitational masses of particles and anti-particles 
are identical or not. It has been argued using a quantum mechanical interpretation of gravitational 
effects that there  might be no difference \cite{gravity_doubters}. However, this should be considered 
premature, because a combination of quantum mechanics and gravity has no solid foundation at present.

There are various experiments which aim to measure antigravity of antiprotonic atoms.
A first step has been made by the ALPHA collaboration at CERN \cite{ALPHA_gravity}. 
They have investigated the time dependence of the decay pattern of antihydrogen. 
It can be excluded that antihydrogen  falls upwards with a gravitational mass $> 65$ 
times its inertial mass \cite{ALPHA_gravity}.
Although this result is not surprising yet, the experiment demonstrates that within the foreseeable future
we can expect a definite experimental answer on whether antimatter differs from matter concerning gravity.
The mass of the proton and the antiproton are not only due to the masses of the constituent quarks
and anti-quarks. The gluons may be the same in both particles as far as gravitation is concerned.
Therefore, the question of antimatter gravity is more complex  than just a difference in sign 
for particle and anti-particle masses; one can rather expect a small difference in the masses. 
The proof of different antiparticle gravitational interaction can enable explanations of present unsolved problems (see e.g. \cite{Villata_2012}).
At CERN-AD several promising experiments are underway to scrutinize antimatter gravity,
the ALPHA \cite{Madsen_2014}, AEGIS \cite{Scampoli_2014} and GBAR \cite{Hamilton_2014} experiments.
Such experiments could be extended towards  purely leptonic systems such as muonium \cite{Kaplan_2013}. In this exotic atom 
the assignment of the label matter or antimatter is not unambiguous, because the system consists of a particle and an antiparticle which have  different masses.

\section{Conclusions}
The dominance of matter over antimatter in the universe surrounding us    
provides a scientific puzzle. First,other than symmetric starting conditions at the big bang cannot be rigorously excluded. Beyond that there are numerous attempts on the theory side next to a number of well motivated experimental approaches aiming at a resolution of at least some of the aspects of the problem. An experimental unambiguous answer can be expected to the question whether there is a gravitational difference between matter and antimatter with a far better than unity precision from several ongoing antihydrogen experiments. This near future result can be foreseen to exclude some of the still open options. For a further going answer we need the results of many different experiments or rather different than present theoretical approaches for significant progress, as no single experiment at this moment is in sight that could give a definitive answer. Therefore explanation of the matter-antimatter asymmetry in the cosmos remains to be one of the biggest scientific challenges.

\bigskip
This work has been supported by the Dutch Stichting voor Fundamenteel Onderzoek der Materie (FOM) in the framework 
of the research programmes 114  ({\it TRI$\mu$P}) and 125 ({\it Broken Mirrors and Drifting Constants}).

\end{document}